\documentclass[a4paper]{cjpsuppl}
\usepackage{write-up}[2002/10/18]
\title{QCD EVOLUTION OF TRANSVERSITY\\ IN LEADING AND NEXT-TO-LEADING ORDER}%
\authori{Philip G. Ratcliffe}%
\addressi{%
  Dip.to di Scienze CC.FF.MM.
\\
  Univ. degli Studi dell'Insubria---sede di Como%
  \,\footnote{The \emph{Insubri} were a Celtic tribe originally from across the
              Alps, who in the 5th.\ century B.C. settled roughly the area now
              known as Lombardy.}
\\
  via Valleggio 11, 22100 Como, Italy
}
\authorii{}    \addressii{}
\authoriii{}   \addressiii{}
\authoriv{}    \addressiv{}
\authorv{}     \addressv{}
\authorvi{}    \addressvi{}
\headtitle{QCD Evolution of Transversity \ldots}
\headauthor{Philip G. Ratcliffe}
\lastevenhead{Philip G. Ratcliffe: QCD Evolution of Transversity \ldots}
\refnum{}
\daterec{20 October 2002;\\final version 31 December 2003}
\suppl{A}  \year{2003} 
\begin{document}
\maketitle
\begin{abstract}
I shall present a rather pedagogical discussion of the transversity
distributions in the quark-parton model and, in particular, the r{\^o}le of
perturbative \acs{QCD} corrections. Among the topics I shall discuss are:
\acs{LO} and \acs{NLO} evolution, the Soffer bound and so-called $K$ factors in
the Drell--Yan process. The main conclusion will be that, compared to
unpolarised or even longitudinally polarised hadron scattering, the case of
transverse spin should actually provide a far clearer window onto the workings
of \acs{QCD} and the interplay with the \acl{QPM}.
\end{abstract}
\section{Introduction}

\subsection{Overview of the talk}

In the past there has been the rather damning prejudice that all
transverse-spin effects (if not indeed spin effects \emph{tout court}) should
\emph{vanish} at very high energies (\ie, where mass effects may be neglected).
This has led to a general lack of interest in the subject on both the
experimental and theoretical sides, with some notable exceptions. This is now
known \emph{not} to be the case.

Indeed, in the near future (and already to some extent) the interest at the
level of the \acr{QPM} in generic deeply-inelastic hadron scattering is due to
shift from unpolarised (and even longitudinally polarised) hadrons to
transversely polarised states. While, on the one hand, the first natural
question to ask is simply the magnitude of the relevant partonic densities, on
the other (however, intimately related), there is the problem of evolution and
the general framework of perturbative \acr{QCD}.

A schematic overview of this talk is then as follows:
{\parskip0pt\itemsep0pt
\begin{itemize} \itemsep0pt\parskip0pt
\item brief history and notation
\item operator-product expansion and renormalisation group
\item \acs{QCD} evolution
{\parskip-\baselineskip
\begin{itemize} \itemsep0pt\parskip0pt
\item leading order
\item next-to-leading order
\item effects on asymmetries
\item effects on the Soffer bound
\vspace*{-0.3\baselineskip}
\end{itemize}
}
\item a \acs{DIS} definition
\item \acs{DIS}--\acs{DY} $K$ factor
\item comments and concluding remarks
\end{itemize}
}

\subsection{A brief history of transversity}

The history of transversity (the concept though not the precise terminology)
begins as early as \citeyear{Ralston:1979ys} with its introduction by
\citet*{Ralston:1979ys} via the Drell--Yan process. Shortly following this the
\acr{LO} anomalous dimensions were first calculated by
\citet*{Baldracchini:1981uq} and \dots promptly forgotten! This decided lack of
interest may be partly traced to the inaccessibility of transversity via the
archetypal parton-model process: namely, \acr{DIS}. Indeed, as we shall see,
the typical process in which transversity may be measured involves at least
\emph{two} polarised hadrons.

A further obstacle was created by the common theoretical prejudice, already
mentioned, according to which precisely \emph{transverse-spin} effects (\ie,
asymmetries) should actually vanish at high energies. The reasons for such a
belief lie in the requirement of chirality-flip in the relevant amplitudes, a
property \emph{not} enjoyed under typical circumstances by a theory of nearly
massless fermions interacting via gauge bosons; however, as shown by
\citet*{Ralston:1979ys}, it turns out that there are indeed several (otherwise
standard) processes in which such effects are on a par with the unpolarised and
helicity-weighted cross-sections.

During the period of great revival witnessed by the spin community, following
the EMC revelations regarding the proton spin, the \acr{LO} anomalous
dimensions for transversity distributions were recalculated by
\citet*{Artru:1990zv}. It is worth recalling that, in fact, these calculations
had also already been, so to speak, unwittingly performed (as contributions to
the evolution of the \acr{DIS} structure function $g_2$) by:
\citet*{Kodaira:1979ib, Antoniadis:1981dg, Bukhvostov:1983te}, and
\citet*{Ratcliffe:1986mp}.

With the typical precision of modern \acr{DIS} measurements, a complete
knowledge of the radiative corrections up to \acr{NLO} is indispensable; in the
case of transversity the \acr{NLO} anomalous dimensions were calculated by:
\citet*{Hayashigaki:1997dn, Kumano:1997qp}, and \citet*{Vogelsang:1998ak}.
Armed with results of such calculations, it is then possible to proceed with an
examination of the phenomenological effects of \acs{QCD} evolution: studies
have been performed by a number of authors; the interested reader is referred
to a recent review paper by \citeauthor*{Barone:2001sp}, where indeed more
details of much of what follows may be found. The lectures by
\citet*{Jaffe:1996zw} also provide a useful pedagogical presentation while an
important early technical discussion laying down the ground rules was given by
\citet*{Jaffe:1992ra}.

\subsection{Notation}

Unfortunately, owing to the somewhat sparse theoretical effort, the literature
now abounds with conflicting notation in regard of the transversity
distributions. For a list and discussion, see Ref.~\cite{Barone:2001sp}, in
accordance with which I shall adopt the form $\DT{f}$ to indicate the
transverse-spin weighted quark density:
\begin{equation}
  \DT{f}(x) = f_\uparrow(x) - f_\downarrow(x) \, ,
\end{equation}
where $f_{\uparrow,\downarrow}(x)$ indicates a parton of type $f$ with
transverse spin vector $\uparrow$ parallel or antiparallel to that of the
parent hadron.

At this point it is worth underlining the fact that while one normally talks of
partonic densities and \acs{DIS} structure functions completely
interchangeably, in the case of transversity there \emph{no} \acs{DIS}
structure function. Thus, any reference to $h_1$ should only be taken as a
generic indication of transversity dependence, with no particular relation to
\acs{DIS}.

\section{Technical Basis}

\subsection{Transverse spin projectors}

Since we are necessarily dealing with transverse spin, it is useful to define
the corresponding polarisation projectors. The transverse polarisation
projectors along the $x$ and $y$ directions (motion is always understood to be
along the $z$-axis) are
\begin{equation}
  \begin{split}
    \mathcal{P}_{\uparrow\downarrow}^{(x)}
    &= \half \, (1 \pm \gamma^1\gamma_5) \, ,
  \\
    \mathcal{P}_{\uparrow\downarrow}^{(y)}
    &= \half \, (1 \pm \gamma^2\gamma_5) \, ,
  \end{split}
\end{equation}
for {positive-energy states} and
\begin{equation}
  \begin{split}
    \mathcal{P}_{\uparrow\downarrow}^{(x)}
    &= \half \, (1 \mp \gamma^1\gamma_5) \, ,
  \\
    \mathcal{P}_{\uparrow\downarrow}^{(y)}
    &= \half \, (1 \mp \gamma^2\gamma_5) \, ,
  \end{split}
\end{equation}
for {negative-energy states}

\subsection{Basis states and amplitudes}

A transversity or transverse-spin basis (with the spin vector $\uparrow$
directed along $y$, for instance) may be expressed in terms of the more
familiar helicity states as
\begin{equation}
  \begin{array}{lcr}
  \ket{\uparrow}   &=& \frac1{\sqrtno2}
  \left[ \, \ket{+} + \I \ket{-} \strut \right] ,
\\[1ex]
  \ket{\downarrow} &=& \frac1{\sqrtno2}
  \left[ \, \ket{+} - \I \ket{-} \strut \right] .
  \end{array}
\end{equation}
The transverse polarisation distributions $\DT{f}$ is then related to an
amplitude that is diagonal in transverse-spin space, while in an helicity base
it is described as an interference effect:
\begin{equation}
  \DT{f}(x) =
  f_\uparrow(x) - f_\downarrow(x) \sim
  \Im \mathcal{A}_{+-,-+} \, .
\end{equation}

\subsection{Chirality flip}

That helicity (or chirality---the terms coincide for massless states) is
flipped in the amplitudes involved is represented pictorially in
Fig.~\ref{fig:chirality}:
\begin{figure}
  \centering
  \includegraphics[width=48mm,bb=158 571 310 681,clip]
                  {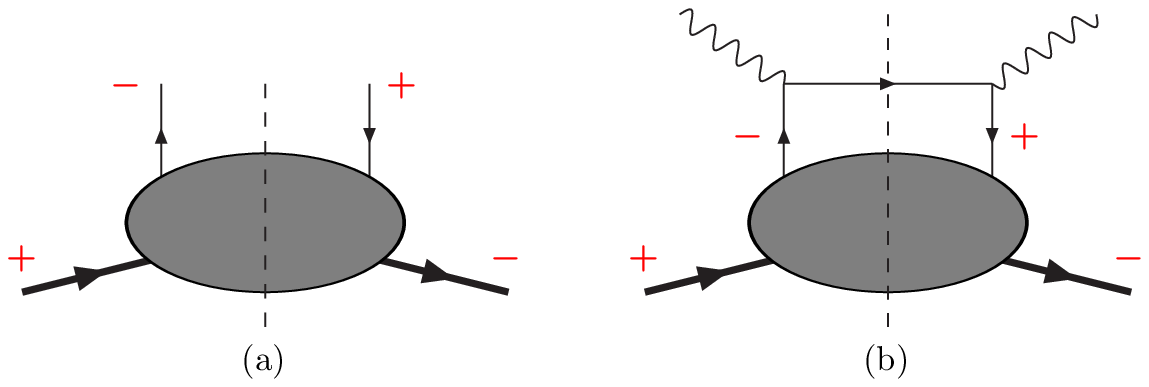}
  \hfil
  \includegraphics[width=48mm,bb=336 571 486 679,clip]
                  {figures/chirality.eps}
  \caption{%
    (a) A chirally-odd hadron--quark amplitude for the $h_1$ partonic density.
    (b) The chirality-flip \emph{forbidden} \acs{DIS} handbag diagram.
  }
  \label{fig:chirality}
\end{figure}
The chirally-odd hadron--quark amplitude contributing to a would-be \acr{DIS}
transversity structure function $h_1$ is depicted in Fig.~\ref{fig:chirality}a.
However, the full \acr{DIS} handbag diagram shown in Fig.~\ref{fig:chirality}b
demonstrates the absence of such a structure owing to the presence of massless
propagators and to helicity conservation at the vector vertices (typical of
gauge theories such as QED and \acs{QCD}). Note, however, that chirality flip
is not a problem if the quark lines of opposite chirality connect to different
hadrons, as for example in the \acr{DY} process.

\subsection{Twist basics and operators}

Let us now place transversity in its proper context, together with the better
known spin-averaged and helicity-weighted parton densities. Note, of course,
although we have just seen that a transversity contribution to fully inclusive
\acr{DIS} is precluded, this is merely due to the nature of that particular
process and not to any fundamental suppression or absence of transversity
itself. Thus, it is more useful to simply consider the corresponding partonic
densities.

Transversity is one of \emph{the} three leading-twist (twist-two) structures:
\begin{eqnarray}
  f(x) &=&
  \int \! \frac{\D\xi^-}{4\pi} \; \E^{\I xP^+ \xi^-}
  \bra{PS}
    \anti\psi(0)
      \gamma^+
    \psi(0, \xi^-, \Vec{0}_\perp)
  \ket{PS} \, ,
\\
  \DL{f}(x)
  &=&
  \int \! \frac{\D\xi^-}{4\pi} \; \E^{\I xP^+ \xi^-}
  \bra{PS}
    \anti\psi(0)
      \gamma^+ \gamma_5
    \psi(0, \xi^-, \Vec{0}_\perp)
  \ket{PS} \, ,
\\
  \DT{f}(x)
  &=&
  \int \! \frac{\D\xi^-}{4\pi} \; \E^{\I xP^+ \xi^-}
  \bra{PS}
    \anti\psi(0)
      \gamma^+ \gamma^1 \gamma_5
    \psi(0, \xi^-, \Vec{0}_\perp)
  \ket{PS} \, ,
\end{eqnarray}
where the state $\ket{PS}$ represents a baryon of four-momentum $P$ and spin
four-vector $S$. The $\gamma_5$ matrix appearing in the second and third lines
signals spin dependence while the extra $\gamma^1$ matrix in $\DT{f}(x)$ signals
the helicity-flip that precludes transversity contributions in \acr{DIS}.

\subsection{Gluon transversity}

Before continuing with a discussion of the quark case, it is worth noting that
transversity may also exist for gluons: it corresponds to linearly polarised
states in a transversely polarised hadron. However, conventional wisdom has it
that, owing to $t$-channel helicity conservation, a spin-half baryon cannot
support the two units of spin-flip necessary for gluon transversity and thus
one is led to the perhaps somewhat surprising conclusion that gluons may
\emph{not} be transversely polarised inside transversely polarised
baryons!\footnote{An interesting case where it might then appear is obviously
the \emph{deuteron.}}

Now, I should point out that such an argument does not take into account
orbital-angular momentum! Let me simply recall that the \acr{AP} kernels
inevitably generate orbital-angular momentum \cite{Ratcliffe:1987dp}; thus it
might be that gluon transversity can be generated in a composite object such as
a baryon; no calculations to such effect exist though. This problem apart, it
is certainly true, as we shall see shortly, that the \emph{quark} and
\emph{gluon} transversity densities evolve \emph{independently}. This fact
alone renders transversity an interesting case for evolution studies---the
subject to which I now turn.

\subsection{The \acs{OPE} and \acs{RGE}}

\begin{figure}
  \centering
  \vspace*{2mm}
  \begin{fmffile}{handbag0}
    \begin{fmfgraph*}(40,30)
      \fmfpen{thick}
      \fmfleft{i1,i2}
      \fmfright{o1,o2}
      \fmf{fermion,tension=0.5}{i1,v1}
      \fmf{fermion,tension=0.5}{v1,v2}
      \fmf{fermion,tension=0.5}{v2,o1}
      \fmf{photon}{i2,v1}
      \fmf{photon}{v2,o2}
    \end{fmfgraph*}
  \end{fmffile}
  \raisebox{15mm}{\quad = \quad $\displaystyle\sum_n C_n$}
  \raisebox{-2mm}{%
    \begin{fmffile}{operator0}
      \begin{fmfgraph*}(20,30)
        \fmfpen{thick}
        \fmfbottom{i1,o1}
        \fmftop{v1}
        \fmf{fermion,tension=0.5}{i1,v1}
        \fmf{fermion,tension=0.5}{v1,o1}
        \fmfv{decor.shape=circle,decor.filled=empty,decor.size=10}{v1}
      \end{fmfgraph*}
    \end{fmffile}
  }%
  \caption{%
    A pictorial representation of the \acs{OPE} as applied to \acs{DIS}.
  }
  \label{fig:OPE-in-DIS}
\end{figure}
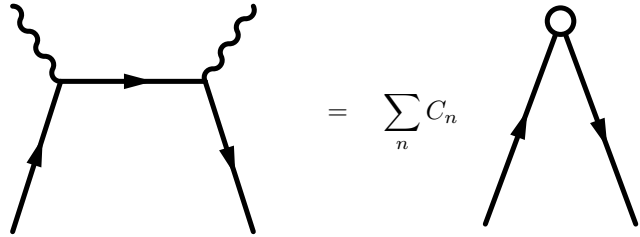

The \acs{OPE}, as applied to \acs{DIS}, is illustrated pictorially in
Fig.~\ref{fig:OPE-in-DIS}.
The anomalous dimensions, $\gamma_n$, are then obtained from the logarithmic
terms in the loop corrections to the right-hand side (\ie, the renormalisation
of the operators $\Oper_n$) while the Wilson coefficients, $C_n$, receive
corrections calculated from the loop corrections to the left-hand side (\ie,
the renormalisation of the hard-scattering cross-section $\hat\sigma$). The
so-formed \acr{RGE} takes the form
\begin{equation}
  \frac{\partial{\Oper_n}(\mu^2)}{\partial\ln\mu^2} +
  \gamma_n \left(\alpha_s(\mu^2)\right) \Oper_n(\mu^2)
  = 0 \, ,
\end{equation}
with standard formal solution
\begin{equation}
  \Oper_n(Q^2) =
  \Oper_n(\mu^2) \,
  \exp
  \left[
    - \int_{\alpha_s(\mu^2)}^{\alpha_s(Q^2)} \!
    \D{\alpha_s} \frac{{\gamma_n}(\alpha_s)}{\beta(\alpha_s)}
  \right] .
\end{equation}

\subsection{Ladder diagram summation}

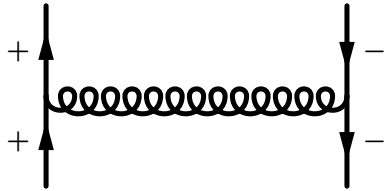
\begin{figure}
  \centering
  \begin{fmffile}{gluonrung1}
    \begin{fmfgraph*}(40,30)
      \fmfpen{thick}
      \fmfset{curly_len}{3mm}
      \fmfbottom{i1,o2}
      \fmftop{o1,i2}
      \fmf{fermion,label=\boldmath{$+$},l.side=left}{i1,v1}
      \fmf{fermion,label=\boldmath{$+$},l.side=left}{v1,o1}
      \fmf{fermion,label=\boldmath{$-$},l.side=left}{i2,v2}
      \fmf{fermion,label=\boldmath{$-$},l.side=left}{v2,o2}
      \fmffreeze
      \fmf{gluon}{v1,v2}
    \end{fmfgraph*}
  \end{fmffile}
  \caption{%
    The \acs{OPI} universal kernel in a physical (axial) gauge governing the
    \acs{QCD} evolution of the partonic transversity distributions.
  }
  \label{fig:1PI-kernel}
\end{figure}

It is instructive to examine the question of evolution within the framework of
the ladder-diagram summation technique \cite{Craigie:1980fa,
Dokshitzer:1978hw}. Recall that the principal tool of this approach is the use
of a physical (axial or light-like) gauge, in which none but the ladder
(planar) diagrams survive the requirement of a large logarithm. In such a gauge
the one-loop \acr{AP} \acr{OPI} kernels for the leading-twist structures are
given by the diagram shown in Fig.~\ref{fig:1PI-kernel}.
In the case of transversity the diagram has a different helicity structure to
those of the spin-averaged and helicity-weighted cases and thus, not
surprisingly, the anomalous dimensions are different in this case.

\begin{figure}
  \centering
  \begin{fmffile}{fermionrung1}
    \begin{fmfgraph*}(40,30)
      \fmfpen{thick}
      \fmfbottom{i1,o2}
      \fmftop{o1,i2}
      \fmf{fermion}{i1,v1}
      \fmf{gluon}{v1,o1}
      \fmf{gluon}{i2,v2}
      \fmf{fermion}{v2,o2}
      \fmffreeze
      \fmf{fermion,label=\boldmath{$+$},l.side=left}{i1,v1}
      \fmf{fermion,label=\boldmath{$-$},l.side=left}{v2,o2}
      \fmf{fermion,label={?}}{v1,v2}
    \end{fmfgraph*}
  \end{fmffile}
  \caption{%
    The \emph{disallowed} \acs{OPI} kernel that would mix the quark and gluon
    contributions to transversity were it not for helicity conservation at the
    vertices.
  }
  \label{fig:1PI-kernel-mixing}
\end{figure}
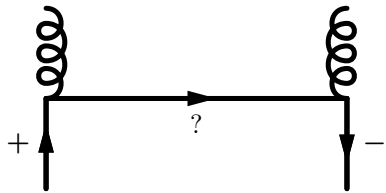

Consider now one of the \acr{OPI} kernels to be calculated for the full
flavour-singlet evolution and that would mix quark and gluon contributions, as
shown in Fig.~\ref{fig:1PI-kernel-mixing}.
Once again the helicity-conserving nature of gauge theories in the massless (or
high-energy) limit leads to a peculiarity in the case of transversity: \acr{LO}
\acs{QCD} evolution of transversity is \emph{non-singlet} like. Thus, even
where a gluon transversity may exist (\eg, in the deuteron) there is no mixing
between the flavour-singlet quark and gluon transversity densities: the two
evolve independently. This means that, for example, for equal statistical
precision, the experimental study of transversity evolution would provide a far
better evaluation of, say, $\alpha_s$; recall that in the spin-averaged and too
in the helicity-weighted cases the strong correlation between $\alpha_s$ and
the ill-determined gluon distributions drastically reduces the significance of
the extracted value of $\alpha_s$. Note also that the usually quoted \acr{DIS}
values for $\alpha_s$ essentially come from sum-rule measurements and thus from
Wilson coefficient corrections and not evolution.

\subsection{Interpolating currents}

\begin{figure}
  \centering
  \begin{fmffile}{higgs0}
    \begin{fmfgraph*}(40,30)
      \fmfpen{thick}
      \fmfleft{i1,i2}
      \fmfright{o1,o2}
      \fmf{fermion,tension=0.5,label={\boldmath{$+$}},l.side=left}{i1,v1}
      \fmf{fermion,tension=0.5,label={\boldmath{$+$}},l.side=left}{v1,v2}
      \fmf{fermion,tension=0.5,label={\boldmath{$-$}},l.side=left}{v2,o1}
      \fmf{photon}{i2,v1}
      \fmf{scalar}{v2,o2}
    \end{fmfgraph*}
  \end{fmffile}
  \caption{%
    A hypothetical Higgs--photon interference term that could contribute to the
    \acs{DIS} cross-section in the case of a transversely polarised hadron.
  }
  \label{fig:DIS-Higgs}
\end{figure}
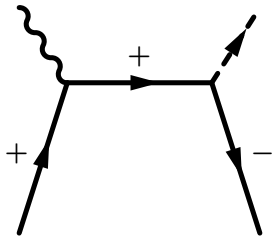

It is also interesting to examine the problem via a method suggested by
\citet*{Ioffe:1995aa}. The idea is simply to use a pair of interpolating
currents that have the correct chirality structure---in this case one vector
and one scalar, with the scalar providing the necessary spin-flip, see
Fig.~\ref{fig:DIS-Higgs}.
The anomalous dimensions are then obtained from the leading logarithmic
corrections to the diagram in Fig.~\ref{fig:1PI-kernel-mixing}. A first attempt
at calculating $\gamma_n$ with this method gave an apparently contradictory
result---subsequently corrected by \citet*{Blumlein:2001ca}. The critical
observation is that while the vector current $J_V$ is conserved and therefore
has $\gamma_V=0$, the scalar current $J_S$ is \emph{not} conserved and thus has
$\gamma_S{\null\not=\null}0$.

Now, the product of two currents may be expanded as
\begin{equation}
  J_V(z) \cdot J_S(0) = \sum_n \, C(n;z) \, \Oper(n;0) \, ,
\end{equation}
and the \acr{RGE} for the Wilson coefficients $C(n;z)$ is
\begin{equation}
  \left[
    \mathcal{D} +
    \gamma_{J_V}(g) +
    \gamma_{J_S}(g) -
    \gamma_{\Oper}(n;g)
    \Strut
  \right]
  C(n;z)
  = 0 \, ,
\end{equation}
where the \acr{RG} operator is
\begin{equation}
  \mathcal{D} =
  \mu^2 \frac{\partial}{\partial\mu^2} +
  \beta(g) \frac{\partial}{\partial{g}} \, .
\end{equation}
Therefore, this chirally-odd interference version of the ``Compton'' amplitude
correction has renormalisation coefficient
\begin{equation}
  \gamma_C(n;g) =
  \gamma_{J_V}(g) +
  \gamma_{J_S}(g) -
  \gamma_{\Oper}(n;g) \, .
\end{equation}
As explained above, while $\gamma_{J_V}=0$ (corresponding to the conservation
of the vector current), $\gamma_{J_S}\not=0$ (the scalar current is \emph{not}
conserved).

We shall discuss later on how this approach suggests a method of examining the
possible $K$ factors involved in the corresponding \acr{DY} process.

\section{\acs{QCD} Evolution}

First, let us now examine a little more thoroughly the evolution problem in
\acr{QCD}, with particular attention to the case of transversity. I shall
discuss the \acr{LO} results in some detail and then simply limit myself to a
demonstration of the effect of including \acr{NLO} corrections.

\subsection{Leading order quark--quark kernels}

The well-known results for the \acr{LO} (indicated by the $0$ index below)
\acr{AP} quark--quark splitting functions in the three twist-two cases are:
\begin{eqnarray}
  P_{qq}^{(0)}
  &=&
  \CF \left( \frac{1+x^2}{1-x} \right)_+ \, ,
\\[2ex]
  \DL{P}_{qq}^{(0)}
  &=&
  P_{qq}^{(0)}
  \quad
  \mbox{helicity conservation,}
\\[1.5ex]
  \DT{P}_{qq}^{(0)}
  &=&
  \CF \left[ \left( \frac{1+x^2}{1-x} \right)_+ - 1 + x \right]
\\[1.5ex]
  &=&
  P_{qq}^{(0)}(x) - \CF (1-x) \, .
\end{eqnarray}
It is useful to define Mellin moments of all quantities involved (partonic
densities, splitting kernels and Wilson coefficients):
\begin{equation}
  f^{(n)} \equiv \int_0^1 \D{x} \, x^{n-1} \, f(x) \, .
\end{equation}
The first moments (\ie, with $n=1$) of the partonic densities often correspond
to sum rules (deriving from conserved quantities or symmetries), which may be
determined independently by other experimental measurements.

Note that for both $P_{qq}^{(0)}$ and $\DL{P}_{qq}^{(0)}$ the first moments
\emph{vanish} (a consequence of vector and axial-vector conservation implying
the existence of sum rules corresponding, \eg, to the total charge and the
neutron beta-decay axial coupling $g_A$) while for $\DT{P}_{qq}^{(0)}$ the same
is not true and the sign implies a falling first moment (the so-called tensor
charge) for transversity. While such a suppression of transversity has obvious
negative implications for high-energy measurements in terms of the size of
effect (asymmetry) one might hope to measure, it does also indicate a more
rapid evolution than in the other two leading-twist cases. This, coupled to the
independence from the gluon density, would imply a greater sensitivity to, for
example, the value of $\alpha_s$.

\subsection{Leading order gluon--gluon kernels}

For completeness, let us now briefly list the corresponding results for the
purely gluonic sector. The three \acr{DGLAP} gluon--gluon splitting functions
at \acr{LO} are as follows:
\begin{eqnarray}
  P_{gg}^{(0)}
  &=&
  \CG
  \left[
    \frac{2x}{(1-x)_+} + 2x(1-x) + \frac{2(1-x)}{x}
  \right]
  \nonumber
\\
  && \hspace{6.5em} \null +
  \left[ \frac{11}6 \CG - \frac23\TF \right] \, \delta(x-1) \, ,
\\[2ex]
  \DL{P}_{gg}^{(0)}
  &=&
  \CG \left[ \frac{2x}{(1-x)_+} + 4(1-x) \right]
  \nonumber
\\
  && \hspace{6.5em} \null +
  \left[ \frac{11}6 \CG - \frac23\TF \right] \, \delta(x-1) \, ,
\\[1.5ex]
  \DT{P}_{gg}^{(0)}
  &=&
  \CG \left[ \frac{2x}{(1-x)_+} \right] +
  \left[ \frac{11}6 \CG - \frac23\TF \right] \, \delta(x-1) \, .
\end{eqnarray}
The first moment in the helicity case, is precisely the leading-order
$\beta$-function coefficient $\beta_0$. Thus, the first moment of the helicity
density $\DL{g}$ grows as $1/\alpha_s$, for the transversity density case
$\DT{g}$ grows \emph{less}, while $g$ (which, of course, is actually infinite)
grows \emph{more} (as $1/x$). All three kernels behave similarly for $x\to1$.

\subsection{Orbital angular momentum}

It is also natural to ask how the question of orbital angular momentum develops
in the case of transversity. Now, since $\DT{q}$ and $\DT{g}$ evolve
independently (recall there is no mixing), the total spin fraction of each of
the two parton types must be conserved \emph{separately}. Thus, in the usual
way, the splitting functions necessarily generate compensating orbital angular
momentum, but for each separately.

Given that $\DT{q}$ \emph{decreases} with increasing $Q^2$, $L_T^q$ must
\emph{increase} in magnitude (assuming a ``primordial'' value of zero) with the
\emph{same sign} as the initial quark spin; the final value will however be
limited. In contrast, $\DT{g}$ increases without bound (just as $\DL{g}$);
thus, $L_T^g$ must also increase in magnitude, but with the \emph{opposite
sign} to the initial gluon spin.

\subsection{\acs{LO} evolution}

\begin{figure}
  \centering
  \includegraphics[width=0.6\textwidth]{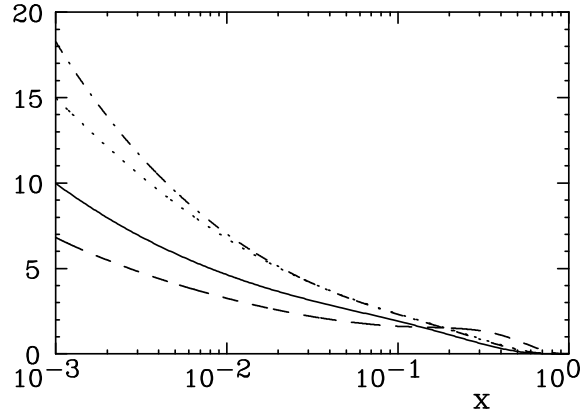}
  \caption{The evolution of $u$-quark helicity and transversity distributions
    compared. The input is $\DT{u}=\DL{u}$ at $Q_0^2=0.23\GeV^2$ (dashed curve).
    The solid (dotted) curve shows $\DT{u}$ ($\DL{u}$) at $Q^2=25\GeV^2$.
    The dot--dashed curve shows the non-singlet evolution of $\DT{u}$ at
    $Q^2=25\GeV^2$ driven by $P_{qq}$.
  }
  \label{fig:evolution-lo}
\end{figure}

The physical implications of evolution for the transversity distributions may
now be examined. Obviously, still lacking is a starting distribution: a
reasonable model may be provided by taking $\DT{q}=\DL{q}$ at some very low
scale. The \acr{LO} evolution for such a hypothetical $u$-quark distribution is
displayed in Fig.~\ref{fig:evolution-lo}.
The relative weakening of the transversity distribution with increasing scale
is evident. The top (dot--dashed) curve shows the evolution of $\DT{u}$
obtained using $P_{qq}$ (in place of $\DT{P}_{qq}$), the difference with
respect to the standard evolution of $\DL{u}$ is due entirely to the lack
(presence) of gluon mixing in the the transversity (helicity) case.

\subsection{Next-to-leading order kernels}

As we move to \acr{NLO} the situation becomes a little more complicated: while
there is still \emph{no} quark--gluon mixing (for the same reasons), there does
arise quark--antiquark mixing due to pair production (as is usual at this
order). In addition, of course, the expressions get much longer and harder to
calculate! The calculations have been performed by three groups:
\citet*{Hayashigaki:1997dn}, \citet*{Kumano:1997qp}, and
\citet*{Vogelsang:1998ak}. In addition, the gluon case has been dealt with by
\citet*{Vogelsang:1998yd}.

The one-loop coefficient functions for \acr{DY} are also known, and in
different renormalisation schemes, see \citet*{Vogelsang:1993jn,
Contogouris:1994ws, Kamal:1996as}, and \citet*{Vogelsang:1998ak}. However, such
corrections are not yet known for any other process.

\subsection{\acs{NLO} evolution}

\begin{figure}
  \centering
  \includegraphics[width=1.0\textwidth]{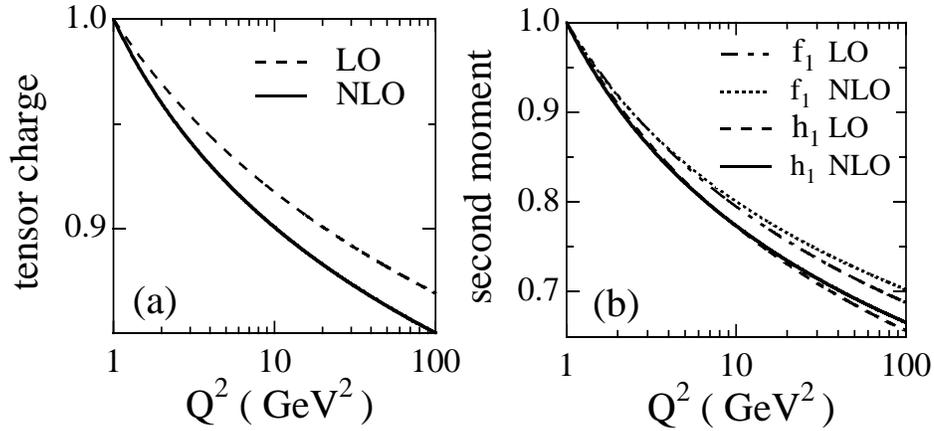}
  \caption{The \acs{LO} and \acs{NLO} $Q^2$-evolution of (a) the tensor charge
    and (b) the second moments of $h_1(x,Q^2)$ and $f_1(x,Q^2)$, all curves are
    normalised to unity at $Q^2=1\GeV^2$ (taken from \cite{Hayashigaki:1997dn}).}
  \label{fig:evolution-nlo-1}
\end{figure}

\begin{figure}
  \centering
  \includegraphics[width=\textwidth]{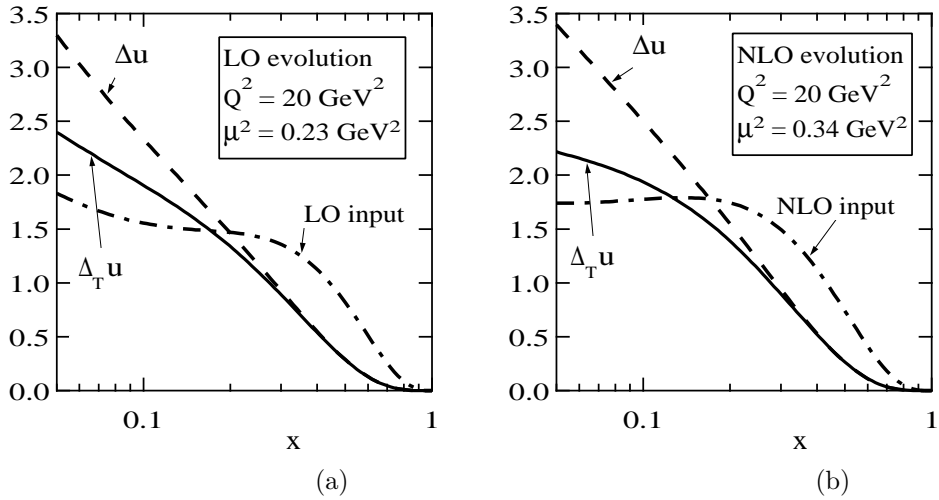}
  \hspace*{3.0cm}(a)\hspace*{6.2cm}(b)
  \caption{A comparison of the $Q^2$-evolution of $\DT{u}(x,Q^2)$ and
    $\DL{u}(x,Q^2)$ at (a) \acs{LO} and (b) \acs{NLO}, assuming the same
    starting values as input (taken from \cite{Hayashigaki:1997dn}).}
  \label{fig:evolution-nlo-2}
\end{figure}

The full \acl{NLO} evolution may thus be studied phenomenologically. Again not
having any data input for $\DT{q}_{qq}$ we must resort to modelling, typically
by assuming equality with the helicity distributions at some starting scale.
The effects of \acl{NLO} evolution on the first two moments are displayed in
Fig.~\ref{fig:evolution-nlo-1};
recall that the vector and axial-vector charges are constant. In
Fig.~\ref{fig:evolution-nlo-2}
a comparison is shown of the effects at \acr{LO} and \acr{NLO}. Note that there
is also a difference in the input moving from \acr{LO} to \acr{NLO} owing to
the differing Wilson coefficients at \acr{NLO}.

\section{The Soffer Bound}

In the case of spin-dependent distributions there exist rather obvious
positivity bounds with respect to the corresponding unpolarised cases: since
the $q_\pm$ are positive definite (at least in the na{\"\i}ve parton model) it
follows that $|\DL{q}|\le{q}$ (the former being the difference and the latter
the sum of the same two positive definite quantities), with an analogous
inequality also holding for $|\DT{q}|$. In addition, the transversity
distribution is constrained by a much less obvious bound derived by
\citet*{Soffer:1995ww}. The derivation, which we shall follow somewhat
schematically, is an instructive example of the necessity of considering
\emph{amplitudes} and not simple \emph{probability densities} when dealing with
problems involving spin states. The central point here is the presence of both
longitudinal \emph{and} transverse spin states; thus either one or the other
must be translated into a different basis.

It is useful to introduce the following hadron--parton amplitudes:
\begin{center}
  \includegraphics[width=0.3\textwidth]{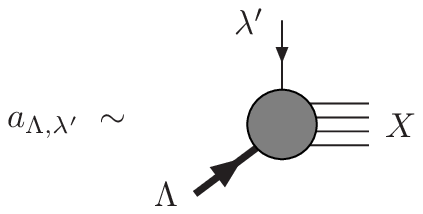}
\end{center}
in terms of which the various partonic densities may be expressed as various
combinations,
\begin{align}
  f(x)
  &\propto
  \Im (\mathcal{A}_{++,++} +\mathcal{A}_{+-,+-})
  && \hspace{-4em} \propto
  \sum_X (a_{++}^* a_{++} + a_{+-}^* a_{+-}) \, ,
  \label{eq:soffer-1}
\\
  \DL{f}(x)
  &\propto
  \Im (\mathcal{A}_{++,++} -\mathcal{A}_{+-,+-})
  &&\hspace{-4em} \propto
  \sum_X ( a_{++}^* a_{++}- a_{+-}^* a_{+-}) \, ,
  \label{eq:soffer-2}
\\
  \DT{f}(x)
  &\propto
  \Im \mathcal{A}_{+-,-+}
  &&\hspace{-4em} \propto
  \sum_X a_{--}^* a_{++} \, .
  \label{eq:soffer-3}
\end{align}
Using these quantities it is then possible to construct a rather non-trivial
Schwartz-type inequality:
\begin{equation}
  \sum_X|a_{++}\pm a_{--}|^2 \ge 0
  \quad \Rightarrow \quad
  \sum_X a_{++}^* a_{++} \pm \sum_X a_{--}^* a_{++} \ge 0 \, ,
\end{equation}
which in turn leads to
\begin{equation}
  f_+(x)\ge|\DT{f}(x)|
  \quad \mbox{or} \quad
  f(x)+\DL{f}(x) \ge 2|\DT{f}(x)| \, .
\end{equation}
This last inequality is precisely the Soffer bound, which interestingly
involves all three leading-twist distributions. Such a bound can, of course,
become particularly stringent in the case of helicity distributions that are
negative, as is the case for the $d$ quark.

\subsection{Evolution of the Soffer bound}

The doubt immediately arises as to whether the bound is respected by \acs{QCD}
evolution; this is not at all a futile question since it is well known that
evolution (in particular, towards \emph{lower} scales) does not even respect
the basic positivity of the \emph{un}-polarised densities. This problem can be
traced to the fact that partonic densities are not physical quantities and thus
beyond the \acr{LO} they are not well defined. A quark seen by a \acr{DIS}
photon may be ``primordial'' in origin (in some definition) or be part of a
$q\bar{q}$ pair created from a primordial gluon (in another). A redefinition of
the densities may lead to a gluonic contribution to the physical \acr{DIS}
cross-section \emph{exceeding} the total cross-section. This will in turn
determine a negative implied value for the primordial quark densities.

Now, the problem is different at \acr{LO} and \acr{NLO}. At leading order there
are no ambiguities and one merely has to inspect the form of the \acr{AP}
kernels. At \acr{NLO} there is no unique definition of the kernels and the
situation is more complicated. Let us start by examining the situation at
\acr{LO}. Maintenance of the Soffer bound under \acs{QCD} evolution has been
argued by \citet*{Bourrely:1997nc}. It is indeed possible to make rather
general arguments: the non-singular terms in the kernels are always positive
definite and thus cannot affect positivity statements. However, the IR singular
(``plus'' regularised) terms in the kernel are negative and thus in principle can
affect inequalities such as that of Soffer. Let us rewrite the plus-regularised
terms in the following manner:
\begin{equation}
  P_+(x,t)
  =
  P(x,t) - \delta(1-x) \int_0^1 \frac{\D{y}}{y} \; P(y,t) \, .
\end{equation}
The \acr{DGLAP} equations can then be recast in a Boltzmann form:
\begin{equation}
  \frac{\D{q}(x,t)}{\D{t}}
  =
  \int_x^1 \frac{\D{y}}{y} \; q(y,t) \, P\!\left( \frac{x}{y},t \right) -
  \int_0^x \frac{\D{y}}{x} \; q(x,t) \, P\!\left( \frac{y}{x\strut},t \right) .
\end{equation}
One sees that the negative term on the right-hand side is ``diagonal'' in $x$ and
thus cannot change the sign of $q(x,t)$, since $q(x,t)$ must go through zero to
turn negative, at which point the evolution switches off. Thus, let us write
\begin{equation}
  \frac{\D{q}_\pm(x,t)}{\D{t}}
  =
    P_{+\pm}(x,t) \otimes q_+(x,t)
  + P_{+\mp}(x,t) \otimes q_-(x,t) \, .
\end{equation}
Then, positivity of the initial distributions, $q_{\pm}(x,t_0)\geq0$ or
$|\DL{q}(x,t_0)|\leq{q}(x,t_0)$, will certainly be preserved if both kernels
$P_{+\PM}$ are positive, which is indeed true. Such an argument can also be
extended in a straight-forward manner to the singlet distributions.

A generalisation of this argument leads to maintenance of the Soffer bound
under \acr{LO} evolution: consideration of the combinations
\begin{equation}
  \label{eq:qcd-super-q}
  Q_\pm(x) = q_+(x) \pm \DT{q}(x) \, ,
\end{equation}
and their evolution kernels indeed demonstrates the stability of the Soffer
bound under \acr{QCD} evolution.

\subsection{Positivity in evolution and \acs{NLO} corrections}

Moving on to \acr{NLO}, as mentioned earlier, the situation is more subtle. A
general comment on positivity constraints concerns the well-known (though
\emph{oft forgotten}) ambiguity in the definition of a partonic density beyond
the \acl{LO} in \acs{QCD}. The physical interpretation of parton distributions
or densities is well-defined and unique in the na{\"\i}ve parton model and in
\acs{QCD} \emph{only} up to the \acr{LLA}. Beyond the \acr{LLA} the coefficient
functions and higher-order \acr{AP} splitting kernels become
renormalisation-scheme dependent. Thus, for some arbitrary scheme adopting a
given starting point (in $Q^2$) where positivity is obeyed, there can be no
guarantee \emph{a priori} of positivity at all $Q^2$.

Such an argument may be turned on its head: that is, such considerations could
provide a criterion for choosing or preferring certain schemes. In other words,
one might decide to adopt only those schemes in which positivity remains
guaranteed at higher orders. However, it should be noted that since the unique
physical meaning of a quark or a gluon beyond the \acr{LLA} is in any case
necessarily lost, such an exercise has probably little or no physical
significance, save perhaps that of possibly endowing numerical evolution
programmes with greater stability. That is, it would avoid the creation of
situations in which there are large (essentially unphysical) cancellations
between opposite sign (and individually positivity violating) polarised quark
and gluon densities---necessary to render the final physical cross-sections
positivity respecting.

\section{A \acs{DIS} Definition for Transversity}

A potentially worrisome and well-known aspect of all phenomenological parton
studies is represented by the presence of non-negligible so-called $K$ factors.
All the other twist-two distribution functions have a natural definition in
\acr{DIS}, where indeed the parton model is usually formulated. However, when
translated to \acr{DY}, for example, large $K$ factors appear in the form of
radiative corrections $\sim\mathrm{O}(\pi\alpha_s)$ to the Wilson coefficients.
At RHIC energies such a correction would be an order 30\% contribution, while
at the lower EMC/SMC energies it could even be as much as around 100\%.

Now, in the case of transversity the pure \acr{DY} coefficient functions are
known to $\mathrm{O}(\alpha_s)$, but are scheme dependent. Moreover, a
$\frac{\ln^2x}{1-x}$ term appears that is not found in either the spin-averaged
or helicity-dependent \acr{DY}. Not only, there is also the problem mentioned
earlier arising in connection with the vector--scalar current product. This
last point is of some relevance as it is connected to a possible (albeit
hypothetical) \acs{DIS}-type process, sensitive to the transversity densities.

\subsection{\acs{DIS} Higgs--photon interference}

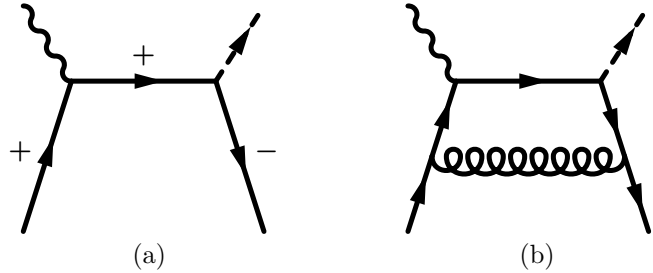
\begin{figure}
  \centering
  \makebox[3cm][c]{%
    \begin{fmffile}{higgs0}
      \begin{fmfgraph*}(40,30)
        \fmfpen{thick}
        \fmfleft{i1,i2}
        \fmfright{o1,o2}
        \fmf{fermion,tension=0.5,label={\boldmath{$+$}},l.side=left}{i1,v1}
        \fmf{fermion,tension=0.5,label={\boldmath{$+$}},l.side=left}{v1,v2}
        \fmf{fermion,tension=0.5,label={\boldmath{$-$}},l.side=left}{v2,o1}
        \fmf{photon}{i2,v1}
        \fmf{scalar}{v2,o2}
      \end{fmfgraph*}
    \end{fmffile}
  }%
  \hspace*{2cm}
  \makebox[3cm][c]{%
    \begin{fmffile}{higgs1}
      \begin{fmfgraph*}(40,30)
        \fmfpen{thick}
        \fmfleft{i1,i2}
        \fmfright{o1,o2}
        \fmf{fermion,tension=1.0}{i1,u1,v1}
        \fmf{fermion,tension=0.5}{v1,v2}
        \fmf{fermion,tension=1.0}{v2,u2,o1}
        \fmf{photon}{i2,v1}
        \fmf{scalar}{v2,o2}
        \fmffreeze
        \fmf{gluon}{u1,u2}
        \fmfv{label=\rnode{NA}{}}{v2}
      \end{fmfgraph*}
    \end{fmffile}
  }
  \hspace*{0.2cm}(a)\hspace*{4.7cm}(b)
  \caption{Higgs--photon interference diagrams: (a) the Born approximation and
    (b) example one-loop contribution to both the \acs{LO} anomalous dimensions
    and the \acs{NLO} Wilson coefficient function.}
  \label{fig:Higgs-gamma}
\end{figure}

In order to obtain a \acr{DIS}-like process in which transversity may play a
r{\^o}le, it is clearly necessary to introduce the possibility of spin-flip. This
essentially means a scalar (or alternatively tensor) vertex. The method of
\citeauthor*{Ioffe:1995aa} effectively has precisely this---a physical
interpretation would be a Higgs--photon interference contribution to the
\acr{DIS} cross-section, see Fig.~\ref{fig:Higgs-gamma}.
The extra logarithmic contribution from the scalar vertex, which was at the
heart of the problem noted earlier, is factorised into the Higgs--quark
coupling constant (or equivalently the running quark mass) and therefore does
not contribute to the \acr{DIS} process.

\subsection[A \acl{DY} $K$ factor]{A \acl{DY} \boldmath{$K$} factor}

Complete evaluation at a numerical level would require inclusion of the full
two-loop anomalous dimensions and the one-loop Wilson coefficient functions.
However, a reasonable first indication may be obtained simply from the one-loop
Wilson coefficient calculated for diagrams such as those in
Fig.~\ref{fig:Higgs-gamma}b. The results are
\begin{subequations}
\begin{eqnarray}
  C^f_{q,\rm DY}-2C^f_{q,\rm DIS} &=&
  \frac{\alpha_s}{2\pi} \, \CF
  \left[
    \frac3{(1-z)_+} + 2(1+z^2)\left(\frac{\ln(1-z)}{1-z}\right)_+
    - 6 - 4z
  \right.
  \nonumber
\\
  && \hspace{10em} \null
    + {\left(\frac43\pi^2+1\right)}\delta(1-z)
  \left.\vphantom{\left(\frac{\ln()}{1}\right)_+} \right] ,
  \label{eq:Wilson-NLO-f}
\\[0.5ex]
  C^g_{q,\rm DY}-2C^g_{q,\rm DIS} &=&
  C^f_{q,\rm DY}-2C^f_{q,\rm DIS} +
  \frac{\alpha_s}{2\pi} \, \CF
  \left[
    2 + 2z
  \right] ,
  \sublabel{subeq:Wilson-NLO-g}
\\[0.5ex]
  C^h_{q,\rm DY}-2C^h_{q,\rm DIS} &=&
  \frac{\alpha_s}{2\pi} \, \CF
  \left[
    \frac{3z}{(1-z)_+} + 4z\left(\frac{\ln(1-z)}{1-z}\right)_+
    + 4(1-z)
  \right.
  \nonumber
\\
  && \hspace{5.3em} \null
    - {6z\frac{\ln^2z}{1-z}}
    + {\left(\frac43\pi^2-1\right)}\delta(1-z)
  \left.\vphantom{\left(\frac{\ln()}{1}\right)_+} \right] ,
  \sublabel{subeq:Wilson-NLO-h}
\end{eqnarray}
\end{subequations}
where $\CF=\frac43$ is the usual colour-group Casimir for the fermion
representation. The three expressions represent the translation coefficient in
going from a \acr{DIS} input to a \acr{DY} output, in other words, quite
literally the difference in the Wilson coefficient relevant to the two cases
(the factor in front of the \acr{DIS} coefficient reflects the fact that two
partons interact in the \acr{DY} process). The first line was first calculated
by \citet*{Altarelli:1978id} and is the correction for the unpolarised
processes, the second is the corresponding correction in the case of
longitudinal polarisation and was first calculated by \citet*{Ratcliffe:1983yj}
and the third expression \cite{Ratcliffe:0000ip} is the corresponding
correction in the case of transversity, using for the \acr{DIS} side the
Higgs--photon interference process described above.

Two substantial differences immediately stand out: firstly, the residues at
$x=1$ are identical in all cases, except for the $\delta$-function
contributions; and secondly, a $\frac{\ln^2x}{1-x}$ term appears in the
transversity case, which is not present in either of the other two cases. This
term actually appears in the \acr{DY} Wilson coefficient and may be traced back
to the different phase-space integration owing to the necessity of \emph{not}
averaging over the azimuthal angle of the final lepton pair.

\begin{figure}
  \centering
  \psfrag{n}{$n$}
  \psfrag{DY-2*DIS coefficient difference}
         {\hspace*{4em}$C^h_{q,\rm DY}-2C^h_{q,\rm DIS}$}
  \includegraphics[width=0.9\textwidth]{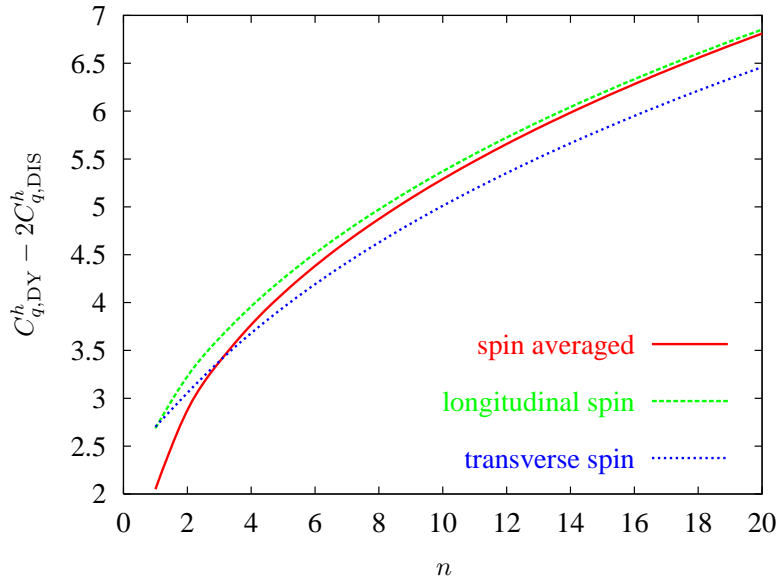}
  \caption{%
    Mellin moments of the \acs{DY}--\acs{DIS} coefficient difference for $q$,
    $\DL{q}$ and $\DT{q}$.
  }
  \label{fig:momdiff}
\end{figure}

By way of comparison, in Fig.~\ref{fig:momdiff}
the moments of the three coefficients, \ie, $q$, $\DL{q}$ and $\DT{q}$ are
shown as a function of moment (recall that higher moments are more sensitive to
larger $x$). Note that while there is convergence between $q$ and $\DL{q}$ for
growing $n$, the transversity coefficient has a rather different behaviour.

\begin{figure}
  \centering
  \psfrag{A_DY}{$A_\mathrm{DY}$}
  \psfrag{tau}{$\tau$}
  \includegraphics[width=0.9\textwidth]{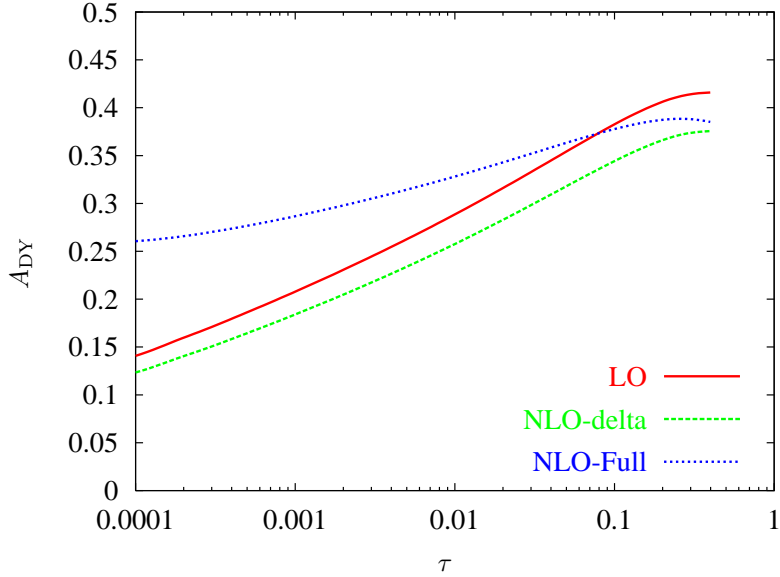}
  \caption{%
    The transversity asymmetry (valence contributions only) for the \acs{DY}
    process. The variable is $\tau=Q^2/s$, with in this
    calculation$s=4{\cdot}10^4\GeV^2$, the kinematical limits are $\tau<x_1,x_2<1$.
  }
  \label{fig:asymdiff}
\end{figure}

The importance of these corrections is best exemplified by an asymmetry
calculation for a physical cross-section. Thus, in Fig.~\ref{fig:asymdiff}
both the \acr{LO} and \acr{NLO} asymmetries are shown for both the helicity and
transversity cases. Note that here only one-loop evolution has been applied;
one would not however expect the two-loop anomalous dimensions to dramatically
alter the effects shown. Again, one sees how the transversity asymmetry differs
substantially from that for helicity (not shown---see \cite{Ratcliffe:1983yj}):
while in the latter case the \acr{NLO} asymmetry slowly converges to the
\acr{LO} calculation for growing $\tau=Q^2/s$ as is to be expected if the large
so-called $\pi^2$ corrections are identical between numerator and denominator
(as indeed is true in the helicity case), in the former the asymmetry
corrections are exceedingly sensitive to variations in $\tau$ and can be quite
large.

Examining the different curves, one sees that there is a non-vanishing
difference for large $\tau$, traceable to the differing residues at $x=1$; and
a still larger difference for small $\tau$, arising from the rather different
functional forms involved in the numerator and denominator. That there should
be such large differences, obviously becoming more important where $\alpha_s$
is larger (\ie, for small $\tau$ and/or $s$), must sound a warning bell to
anyone considering making predictions based on models normalised to \acr{DIS}
distributions, and likewise to anyone wishing to extract densities from
\acr{DY}-like measurements.

At this point one might object that the higher-order splitting kernels have
also now been calculated, indeed for all three cases---see below, and thus the
usual ambiguities are really only present at \acr{NNLO}. In fact, the
calculation of the two-loop anomalous dimensions for $h_1$ has been presented
in three papers: \citet*{Hayashigaki:1997dn} and \citet*{Kumano:1997qp} used
the \acr{MS} scheme in the Feynman gauge while \citet*{Vogelsang:1998ak}
adopted the \acr{MMS} scheme in the \acl{LC} gauge. These complement the
earlier two-loop calculations for the two other better-known twist-two
structure functions: $f_1$ \citep*{Floratos:1977au, Floratos:1979ny,
Gonzalez-Arroyo:1979df, Curci:1980uw, Furmanski:1980cm, Floratos:1981hk,
Floratos:1981hm} and $g_1$ \citep*{Mertig:1996ny, Vogelsang:1996vh,
Vogelsang:1996im}. However, this is not quite the point, indeed there is
actually no ambiguity in the expressions
(\ref{eq:Wilson-NLO-f}--\ref{subeq:Wilson-NLO-h}).

Most model calculations make some (albeit indirect) reference to \acr{DIS} and
transversity densities are then normalised in parallel with the unpolarised
densities. Thus predictions for a \acr{DY} cross-section should, for
consistency, include something like the corrections calculated here. Of course,
it is hard to make the claim that the approach adopted here provides precisely
the form of correction that really applies. However, the fact that even at the
level of an asymmetry large corrections remain must be taken as a warning that
transversity densities too could reserve surprises. Note that such observations
have absolutely no relevance though to the question of pure \acr{QCD}
evolution.

\section{Comments and Concluding Remarks}

By way of concluding remarks let us simply try to recapitulate the important
points touched in this all too brief presentation. First a few well-understood
and theoretically clear points:
\begin{itemize} \itemsep0pt
\item
Both the non-singlet and non-mixing behaviour render transversity surprisingly
simpler and more transparent to study, with respect to its better-known
siblings, both from an experimental and theoretical point of view.
\item
At high energies \acr{QCD} evolution suppresses $\DT{q}$ with respect to both
$\DL{q}$ and ${q}$; thus, first measurements will best be performed at lower
values of $Q^2$. However, complementary high-$Q^2$ measurements will always be
required to perform meaningful evolution studies.
\item
The previous observation may be turned on its head: transversity will be a
wonderful place to study \acr{QCD} evolution as even the first moment evolves
rather rapidly.
\end{itemize}
On the other hand, there are also aspects that appear to be less well
understood and that could therefore well lead to surprises:
\begin{itemize} \itemsep0pt
\item
If the calculations reported here are at all indicative, the well-known large
$K$ factors involved in the translation between \acr{DIS} and \acr{DY} may, in
the case of transversity, lead to rather unstable asymmetries and thus poorly
defined extracted partonic densities.
\item
If the argument leading to the conclusion that gluon transversity is excluded
from spin-half baryons should turn out to be flawed, this might be a new
indication of the importance of orbital angular momentum effects.
\end{itemize}

I should remark that there has been neither the time or space here to discuss
the very rich and interesting phenomenology associated with single-spin
asymmetries, which could also turn out to be related to transversity (see, for
example, \cite{Barone:2001sp} and references therein).

As a final word then, it should now be obvious that transverse-spin effects,
far from being negligible and uninteresting at high energies, already from a
solid theoretical viewpoint actually promise an interesting window onto the
workings of \acr{QCD} evolution. Moreover, the possibility of further
spin-driven surprises from this experimentally new sector is not to be ignored
and the theory community is now eagerly awaiting the first data.


\end{document}